\documentclass{article}
\usepackage[a5paper,tmargin=1.5cm, bmargin=1.5cm, lmargin=1.5cm, rmargin=1.5cm]{geometry}

\usepackage[utf8]{inputenc}
\usepackage[T1]{fontenc}
\usepackage{lmodern}	

\usepackage[english]{babel}
\usepackage{url}
\usepackage{microtype}

\usepackage{listings}
\def\test{1}   
\if\test1\usepackage{xcolor}
\fi

\title{Fortran... ok, and what's next?}

\author{Vincent Magnin\footnote{Lecturer, IEMN (UMR CNRS 8520), vincent.magnin@univ-lille.fr}, José Alves\footnote{Development Team Leader on Multiphysics and Deep-Learning, Transvalor S.A., Sophia Antipolis, jose.alves@transvalor.com}, Antoine Arnoud\footnote{Economist, International Monetary Fund, antoine.arnoud@gmail.com}, Arjen Markus\footnote{Responsible for numerical models dedicated to water quality and ecology, Deltares Research Institute, Pays-Bas, arjen.markus895@gmail.com}, \\ Michele Esposito Marzino\footnote{Doctoral student at the University of Liège, michele.espositomarzino@uliege.be}}
\date{}

\begin{document}

\maketitle

\small
\fontfamily{phv}\selectfont
This is the official translation of the original French paper ``Fortran... et puis quoi encore ?'' published in the \textit{Bulletin 1024} of the Société Informatique de France. This translation was produced in two steps: first, tools based on deep learning were used, then we reviewed the translation line by line and improved it manually. The article can  be cited as:

Magnin V., Alves J., Arnoud A., Markus A., and Esposito Marzino M., ``Fortran... et puis quoi encore ?'', \textit{Bulletin 1024,} no.~22, pp.143--161, November 2023, DOI:10.48556/SIF.1024.22.143, also available in English: ``Fortran... ok, and what's next?''.

{\centering \rule{0.5\columnwidth}{0.4pt}\par}
~
\\
\normalsize
\fontfamily{lmr}\selectfont
\textit{
Modern Fortran is a standardized language that includes object-oriented and parallel programming paradigms. The Fortran-lang community, created at the end of 2019, is actively working to modernize its ecosystem. New compilers are under development. And the fourth Fortran standard of the 21st century is due to be published in autumn 2023 \footnote{Addendum: Fortran 2023 was published by ISO on November 27, 2023. The most recent J3 Interpretation Document is \url{https://j3-fortran.org/doc/year/24/24-007.pdf}.}.
}

\section{Genesis and classical period}

\subsection{The first optimizing compiler}

With his Master's degree in mathematics, the young John Backus was hired by IBM in 1950 to work on the company's calculators. To lighten his machine-language programming load, he first designed an interpreter called Speedcoding for the IBM 701, featuring floating-point emulation. Then, in December 1953, he proposed the FORTRAN\footnote{The name was chosen at the end of 1954, the initial proposal being for an ``automatic coding system''. Note that FORTRAN is written for the period prior to Fortran 90.} \textit{(mathematical FORmula TRANslating system)} project. Machine language programming requires experts, and at least half the cost of a scientific calculation is due to their salaries (programming, testing, debugging). A high-level language would drastically reduce the amount of work required. This economic argument enabled Backus to rapidly obtain a team to work on a language and its compiler \cite{Backus1957}. Everything had yet to be invented. While the idea of a compiled language was already in the air with Grace Hopper's work on the A-0 language, the FORTRAN compiler\footnote{In fact, the articles of the time use the word \textit{translator}, rather than \textit{compiler}.} is considered to be the first optimizing compiler. The team was indeed aware that, for it to be accepted, optimal code had to be generated whatever the application. At the time, few people were prepared to believe that a machine could generate machine code as fast as that written by a human being. And von Neumann himself saw no need for a high-level language \cite{lorenzo_abstracting_2019}.

The target machine was the future IBM 704, and Backus insisted that it should be the first machine to work directly with floating-point reals\footnote{In PCs, it wasn't until 1989 with the Intel i486 32-bit microprocessor that the floating-point arithmetic coprocessor (FPU) was no longer optional.}. After three years' work, the compiler was finally deployed in April 1957. It was a revolution. Mary Tsingou, physicist and mathematician at Los Alamos National Laboratory, describes the situation: ``When Fortran came, it was almost like paradise'' \cite{marytsingou}. Its syntax is indeed simple and adapted to scientific calculations, as its name implies. The programmer's reference manual, ready in October 1956, is 54 pages long \cite{manuel1956}, so the basics of the language can be learned in a few hours. Frances Elizabeth Allen, hired at the age of 24 in July 1957, was in charge of teaching FORTRAN to IBM scientists, and the hardest part was convincing them of its usefulness and efficiency. But reading the compiler's 20,000 lines of code would guide her career. She was awarded the 2006 Turing Award ``for pioneering contributions to the theory and practice of optimizing compiler techniques that laid the foundation for modern optimizing compilers and automatic parallel execution.'' \cite{prixturing2006}

Designed with only the IBM 704 in mind, the FORTRAN compiler quickly proved adaptable to other machines, starting with IBM's own, making it the first portable programming language. The programmer could now abstract from machine language, and the scientist could improvise as a programmer \cite{lorenzo_abstracting_2019}. So it's hardly surprising that, according to an article in the journal Nature published in 2021, this compiler is one of the ten computer codes that have most transformed science \cite{perkel_2021}.

\subsection{The first standardized language}

On the downside, standardization soon became necessary to maintain the language's portability, with each compiler writer adding their own extensions. It therefore became the first language to be standardized with the FORTRAN 66 standard (ANSI X3.9-1966), which includes two versions of the language: the full language and a simplified version called Basic FORTRAN. An intermediate version, ECMA-9, was even standardized a year earlier by the ECMA \textit{(European Computer Manufacturers Association).}

\subsection{FORTRAN 77}

The FORTRAN 77 standard (ANSI X3.9-1978), supplemented by the U.S. Department of Defense's MIL-STD-1753 extension, is a major development that facilitates structured programming in particular. 

The military extension also introduces the \texttt{IMPLICIT NONE} instruction, which is now systematically used, to disable implicit variable typing. Indeed, by default, variables whose names start with I, J, K, L, M or N are of integer type, while all others are of real type. These are the classic loop counters \texttt{i, j, k} that we still use!

The language remained dominant until the late 1970s. So much so, in fact, that many artists used it to explore the new possibilities offered by the computer: for example, to create algorithmic music (Iannis Xenakis and Pierre Barbaud), graphic works (Hiroshi Kawano), poem generators and so on. And FORTRAN 77 would remain a benchmark for a long, long time to come!

\subsection{A Turing Award}

Among other honors, John Backus received the 1977 Turing Award ``for profound, influential, and lasting contributions to the design of practical high-level programming systems, notably through his work on FORTRAN, and for seminal publication of formal procedures for the specification of programming languages.''\cite{prixturing1977} Once the FORTRAN project was completed, he quickly moved on to other adventures. His participation in ALGOL 58 and 60 led him to work on what came to be known as the Backus-Naur Form (BNF), a formal notation for describing a programming language. Appointed \textit{IBM Fellow} in 1963, he went on to work on his vision of what a functional programming style could be, his Turing reading being entitled \textit{Can programming be liberated from the von Neumann style?: a functional style and its algebra of programs.}.

\section{Modern Fortran}

The expression \textit{Modern Fortran}, often used in the titles of Anglo-Saxon books over the last twenty years or so \cite{curcic_modern_2020, markus2012, metcalf_modern_2018}, has nothing official about it: some will use it from Fortran 90/95 onwards, others from Fortran 2003 onwards. It could be perceived as an oxymoron, a figure of speech common in our times, but it's not!

\subsection{Fortran 90}

The Fortran 90 standard, which arrived several years late, is a major revision, as suggested by the change of case from FORTRAN to Fortran. This is also its first ISO standard, managed by the international committee ISO/IEC JTC1/SC22/WG5 and the American technical committee X3J3. Among the many new features is the \textit{free form} source code format, which at last makes it possible to write code without worrying about the role of certain columns inherited from punched cards (\textit{fixed form} format, generally associated with the extension \texttt{.f}). Note that the most commonly used extension for the free format is \texttt{.f90}, which can be confusing for beginners, as the code can conform to any subsequent standard!

Rather than defining new types, the concept of \textit{kind} was introduced to define potential variants of all basic types (\texttt{real}, \texttt{complex}, \texttt{integer}, \texttt{logical}, \texttt{character}). For \texttt{integer} and \texttt{real}, intrinsic functions can be used to select a variant, if supported by the compiler, according to the desired numerical characteristics.

The division of source code into modules enables an embryo of object-oriented programming, with each module embedding its own data and routines, and facilitates compiler diagnostics. The \texttt{public} (default) and \texttt{private} attributes can be used to indicate whether or not an entity will be accessible outside the module. In addition, the concept of derived type allows users to define their own data structures:

\begin{lstlisting}
  type person
    character(30) :: name
    integer :: age
    real :: weight
  end type

  type(person) :: him = person("Jean", 50, 70.5)
  print *, "His name is: ", him%name
\end{lstlisting}

Note the use of the \texttt{\%} character instead of the classic dot to access a field in the structure. The introduction of derived types is all the more important as they will form the basis for the introduction of object-oriented programming in the Fortran 2003 standard.

Array-oriented syntax simplifies code by avoiding numerous loops. In the example below, three vectors with three components are defined \footnote{The strict syntax for initializing an array in Fortran 90 is \texttt{(/ /)}, but Fortran 2003 introduced the more readable and intuitive syntax of square brackets.}, the sum is calculated and displayed, and a vector is calculated whose components are the square roots of the components of the first. Fortran 90 indeed introduced so-called \texttt{elemental} functions and procedures, which can be applied to both scalars and arrays.

\begin{lstlisting}
program arrays
  implicit none
  real :: u(3) = [ 1.0, 2.0, 3.0 ]
  real :: v(3) = [ 4.0, 5.0, 6.0 ]
  real :: w(3)

  w = u + v
  print *, "Sum of vectors u and v: ", w
  print *, "Square roots vector of u(i): ", sqrt(u)
end program arrays
\end{lstlisting}

In addition, arrays can now be dynamically allocated. In this case, they are declared with the attribute \texttt{allocatable} and their dimensions are then defined later using the instruction \texttt{allocate()}. Occupied space can be freed with the instruction \texttt{deallocate()}.

Pointers (attribute \texttt{pointer}) made their appearance in Fortran, facilitating the use of lists, trees, graphs and so on. But whereas in the C language a pointer is a memory address, the Fortran concept of a pointer is more akin to an alias. Moreover, an entity can only be pointed to if it has been declared with the attribute \texttt{target} or if it is itself a pointer. These properties enable compilers to generate robust, optimized code.

Finally, two new sections appear in Fortran standards: the list of obsolescent features and the list of deleted features. In practice, the latter are generally still supported by compilers, so that older code can continue to function without change.

\subsection{Fortran 95}

This is a minor standard that complements the previous one. Fortran 90 had introduced the possibility of allocating memory and using pointers to arrays. But these features had their limits, as the programmer had to manage memory explicitly. Fortran 95 offers much greater freedom and flexibility: automatic deallocation frees the programmer from much of the memory management, limiting memory leakage problems and allowing them to concentrate even more on what they want to do rather than on the technical details of getting there.

A function can be given the attribute \texttt{pure}, indicating that it produces no side effects (modification of a global variable or an argument, input/output realization, etc.) and can therefore be used without problem in a parallel computation. The programmer can also define their own \texttt{elemental} functions, a particular form of \texttt{pure} function.

As a side note, one of the methods used to evolve the language is therefore to add attributes such as \texttt{elemental}, \texttt{pure}, \texttt{intent}, \texttt{allocatable}, \texttt{pointer}, \texttt{target}\dots~They also ensure the robustness of the code, enabling the compiler to perform the necessary checks for each.

Finally, the first attempt to introduce parallel computing into the language itself, \texttt{forall} loops will eventually be declared obsolescent in the Fortran 2018 standard, their complexity having made it difficult for compilers to implement them efficiently.

\subsection{Fortran 2003}

Just like Fortran 90, Fortran 2003 is a major release, as evidenced by its 585 pages, 55 \% more than the previous standard. In particular, it introduces object-oriented programming (OOP) \cite{rouson2014}, based on the modules and derived types introduced by Fortran 90. A class is thus a derived type containing both data and methods, called type-bound procedures, as in this simple example:

\begin{lstlisting}
type :: vector2D
  real :: x, y
contains
  procedure :: add => add_vector2D
  procedure :: rotation => rotation_vector2D
end type vector2D
\end{lstlisting}

The introduction of classes makes true OOP possible: more or less inspired by the concept of class in languages such as C++ and Java, a Fortran class can be abstract (attribute \texttt{abstract}) or not, and can be extended (attribute \texttt{extends}). But one can also customize the methods of objects belonging to the same class by using procedure pointers, the syntax being similar, and thus achieve a prototype-oriented programming style \cite{prototype}.

To further facilitate the construction of abstract data types such as linked lists, we can also use so-called unlimited polymorphic variables. These can be described as chameleonic, as they take on the value assigned to them, while retaining the initial type, thus maintaining safe typing \cite{markus2012}.

Another major novelty concerns the long awaited standardization of the interoperability with the C language. This makes it easy to call C libraries in a portable way, for example to create graphical interfaces with GTK~\cite{gtk-fortran2019} or manage databases with SQLite \cite{markus2012}. Conversely, it makes it easier to write Fortran libraries for other languages, such as Python, enabling them to benefit from its computational speed and numerical precision. In practice, the new intrinsic module \texttt{iso\_c\_binding} contains constants for mapping between the data types of the two languages, as well as functions for mapping between their pointers. As the order in which multi-dimensional arrays are stored in memory is different in the two languages, the programmer must take care to reverse the order of indices when such an array is passed from one language to the other.

Both the \texttt{move\_alloc()} procedure and implicit allocation on assignment simplify the dynamic management of \texttt{allocatable} objects, including reallocation.

Among the many other new features are I/O improvements (asynchronous transfers, for example), three intrinsic modules providing access to IEEE 754 standard functions (e.g. changing the microprocessor rounding mode), optional support for Unicode characters (ISO/IEC 1064619 1:2000 UCS-4), access to command-line arguments and environment variables.

\section{Parallel Computing}

Some Fortran codes are designed to run on personal computers, while others are intended to be run on supercomputers. Parallelism must therefore be managed at different scales and on different types of architectures \cite{calculparallele}: multi-core microprocessors, vectorization \footnote{ SIMD instructions (\textit{single instruction, multiple data}) such as FMA, SSE, AVX, etc. Modern compilers are able to vectorize certain loops automatically, and standards such as OpenMP include SIMD directives enabling the developer to guide the compiler in complex cases.}, graphics processors (GPUs), multi-processor machines of various architectures, networked machines, etc. Setting up parallel computing standards and implementing them efficiently is therefore a difficult task.

The first attempts date back to the second half of the 80s, at a time when supercomputers were evolving from a vector model to a multi-processor model, and network computing was becoming more widespread. In 1989, PVM \textit{(Parallel Virtual Machine)} \cite{PVM} was one of the first solutions for parallel computing in C or Fortran on a network of heterogeneous machines (Unix/Windows).

At the end of 1991, work began on standardizing High Performance Fortran (HPF)\cite{kennedy2007}. With everything still to be invented, this version of the language failed to convince and never reached maturity, except in Japan, which developed its HPF/JA dialect and used it on the Earth Simulator supercomputer, the most powerful in the world from 2002 to 2004 \cite{TOP500}. But HPF helped to lay the foundations, and a number of seeds were sown. For example, the addition of compiler directives in comments to ensure that the program remains compilable on non-parallel machines will be found in OpenMP or OpenACC, as will the \texttt{forall} loops and \texttt{pure} attribute in Fortran 95.

The inclusion of parallel computing paradigms in the Fortran standards themselves would require an enormous amount of time and experimentation. This would have to wait until Fortran 2008 and 2018, with the considerable effort this would entail in compiler development.

\subsection{External Standards}

Two external standards emerged in the 90s for the C, C++ and Fortran languages \cite{kennedy2007}. Released in 1994, MPI \textit{(Message Passing Interface)} enables parallel programming by passing messages between processes. It is designed to work with both shared-memory and distributed-memory systems. The latter may involve computers in a network, nodes in a computing cluster, etc. Powerful, but requiring the user to control inter-process communications, MPI is sometimes referred to as the assembler of parallel computing. Released in 1997, OpenMP \textit{(Open Multi-Processing)} is designed for shared memory systems. Using comment directives, it enables existing code to be parallelized quite easily, if the algorithms are suitable.

The rise of GPU computing power in the 2000s led to the emergence of various systems enabling calculations to be offloaded to the many cores of these processors, including NVIDIA's CUDA \textit{(Compute Unified Device Architecture)} in 2007, OpenCL 1.0 \textit{(Open Computing Language)} in 2009, and OpenACC 2.0 \textit{(Open Accelerators)} in 2012.

But these standards are external to Fortran. The next two standards would introduce parallel computing paradigms into the language itself.


\subsection{Fortran 2008}

Initially developed as an extension to Fortran 95, {\it coarrays} are officially included in the Fortran 2008 standard. 
Several copies of a program can be executed in parallel, either on a multi-core processor or a supercomputer. 
Each copy, called an {\it image}, holds its own data\footnote{\textit{Single program, multiple data (SPMD).}}. 

The intrinsic functions \texttt{num\_images()} and \texttt{this\_image()} provide the number of images and the number of the current image (named {\it co-index}). 
The syntax is simple and inspired by tensorial notations \cite{numrich2005}. 
While Fortran array indexes are within parentheses, the co-index is within brackets. 
Without explicit brackets, reference is made to the co-array of the current image. 
Using brackets, co-arrays of other images are accessed transparently. 
When needed, it is possible to synchronize images using, for example, the {\tt sync all} instruction. 

Here is a simple example to compute an approximation of $\pi$ using a Monte Carlo method 
(an inefficient algorithm but suitable for this example!). 
We declare a co-array {\tt k[*]} of 64-bit integers\footnote{These are scalars to simplify the example, 
but a co-array can also be declared from a classic array.} which is zero initialized in each image 
and incremented when a random point drawn inside a square of side length 2 falls into the inscribed 
circle of unit radius. 
The star within the brackets means that the number of co-dimensions will not be known until execution. 
Once all the images reach the {\tt sync all} instruction, image 1 is responsible for retrieving 
and adding the results of all images, and finally print the approximation of $\pi$.

\begin{lstlisting}
program pi_monte_carlo_coarrays
  use, intrinsic :: iso_fortran_env, only: wp=>real64, int64
  implicit none
  real(wp)       :: xy(2)
  integer(int64) :: i, j, kt=0, k[*]=0
  integer(int64), parameter :: n=1000000000

  do i = 1, n / num_images()
    call random_number(xy)
    if (sum(xy**2) < 1) k = k + 1
  end do

  sync all
  if (this_image() == 1) then
    do j = 1, num_images()
      kt = kt + k[j]
    end do
    print *, (4.0_wp * kt) / n
  end if
end program pi_monte_carlo_coarrays
\end{lstlisting}

Fortran 2008 also introduces {\tt do concurrent} loops which tell the compiler that the code 
inside the loop does not make use of data coming from previous iterations, 
and so it can be parallelized if the compiler considers it beneficial. 
For example the NVIDIA compiler nvfortran is capable of offloading execution of such a loop on their NVIDIA GPUs. 
These loops will eventually replace {\tt forall} parallel loops that Fortran 95 standard tried to introduce. 

Two new features allow for a better code structuring. 
Large modules can be split into {\tt submodules}. 
Also, if originally all declarations had to come before the first executable statement, 
the {\tt block...end block} construct allows for local declaration of variables 
(or module imports) at any point of the code. 
Hence, the code structure of a long procedure can be made more readable. 

Among other additions in the Fortran 2008 standard, we find the intrinsic module \texttt{iso\_fortran\_env} 
which provides standardized {\tt kind} constants clearly identifying the types of data (for example 
{\tt real64} and {\tt int64}), the maximum number of dimensions of an array (named {\it rank}) 
going from 7 to 15, the possibility of launching another program through the command line ({\tt execute\_command\_line}), 
and the {\it contiguous} attribute which tells the compiler that an array received as a procedure argument 
will occupy a contiguous memory area\footnote{Which could potentially improve the performances.}. 
Finally, many intrinsic functions are added, particularly for scientific computation: 
inverse trigonometric functions with complex arguments, inverse hyperbolic functions, Bessel functions, 
error and gamma functions, etc. 

\subsection{Fortran 2018}

Fortran 2018 brings numerous improvements, for example concerning C interoperability, 
but more importantly new concepts for parallel computing.
Images can be grouped into {\it teams} working in parallel on different tasks. 
The concept of {\it events} allows images to communicate easily without requiring co-arrays. And {\it collective subroutines} allow performing, in a simple way, reduction operations such as 
computing the sum of the values of a variable in each image. 
The previous example on the Monte Carlo computation of $\pi$ can thus be rewritten more concisely:

\begin{lstlisting}
program pi_monte_carlo_co_sum
    use, intrinsic :: iso_fortran_env, only: wp=>real64, int64
    implicit none
    real(wp)       :: xy(2)
    integer(int64) :: i, k=0
    integer(int64), parameter :: n=1000000000

    do i = 1, n / num_images()
        call random_number(xy)
        if (sum(xy**2) < 1) k = k + 1
    end do

    call co_sum(k, result_image = 1)
    if (this_image() == 1)  print *, (4.0_wp * k) / n
end program pi_monte_carlo_co_sum
\end{lstlisting}

Intel compilers were the first to fully implement Fortran 2018 and since 2020 they are freely available, 
as all the other {\it Intel oneAPI Toolkits}. 
With the 2023 version, the classic ifort compiler is giving way to the new ifx compiler, 
based on the LLVM compiler infrastructure. 

On the side of open source compilers, GFortran (GCC) is the most used. 
It still does not fully implement Fortran 2018 and relies on the OpenCoarrays library for parallel computing, 
yet it has the essentials and its development progresses. 

\section{Fortran and its communities}

\subsection{From the origins}

Recent programming languages often have an organized community on the internet.
But Fortran did not wait for computer networks to have its communities.
In 1955, IBM created the SHARE user group for its IBM 704 machines.
Backus and his team presented and discussed through this group their work on the language and its first compiler.
The following paragraph describes the situation after its distribution \cite{heising_emergence_1984}:

\begin{quote}
   {\it Nonetheless, the technical basis of FORTRAN was sufficiently sound that usage was like a snowball going downhill. Soon there were hundreds of customers making hundreds of suggestions for improvements. They would find bugs and send them in – not only error reports, but in many cases the fixes would come in along with the reports. Many suggestions applied to such matters as improvement of diagnostics – little practical things – and it was as if there were hundreds of people working on improving FORTRAN. The suggestions just poured in, and we put them in as fast as we could.}
\end{quote}

In 1970 the \textit{Fortran Specialist Group} of the \textit{British Computer Society}\footnote{\url{https://fortran.bcs.org/}} was founded,
which continues today its missions consisting in promoting the language,
working to its evolutions and fostering exchanges and meetings between users.

\subsection{Usenet}

Following the creation of the Usenet network, a discussion group \\
\texttt{net.lang.f77} was created and later renamed \texttt{comp.lang.fortran} in 1986.
It would remain for a long time the main place online where Fortran is discussed.
It is still used, with typically ten new discussion topics per month.
Nowadays, Usenet servers are scarce, including in universities, 
but one can also access these discussion groups via Google Groups.

\subsection{Some reference websites} 

The Fortran Wiki\footnote{\url{http://fortranwiki.org/}}, created in 2008, has become an 
important source of information on the language, by allowing everyone to enrich it.  

The Fortran French Wikipedia page is kept up to date and offers a lot of 
information and references on the language and its ecosystem.

A site\footnote{\url{https://www.fortran90.org/}} created in 2012 by Ondřej Čertík will long remain a 
reference in the Fortran world. Its content began to be transferred to the Fortran-lang community site.

\subsection{The Fortran-lang community} 

Despite all these tools, it must be recognized that the galaxy of Fortran users remained 
rather scattered and poorly organized (except in standardization committees), 
compared to more recent languages. 
For a long time, there were many researchers or engineers publishing 
their libraries and programs in isolation on their websites.

At the end of 2019, a handful of developers decided to remedy this situation.
They came together and created the Fortran-lang community using the GitHub platform to centralize developments \cite{kedward_state_2022}.
They thus created a central site\footnote{\url{https://fortran-lang.org/}} to provide all the information and resources needed by both the beginner and the expert programmer alike.
To federate the community, they also set up a modern forum, the Fortran Discourse\footnote{\url{https://fortran-lang.discourse.group/}}, 
which now has over 1,100 registered users (from students to Fortran Standards Committee members).
The most involved in the organization meet to discuss at monthly videoconferences\footnote{The community 
began to grow at the time of the first COVID lockdowns.}

\subsubsection{The stdlib standard library} 

Having made a diagnosis of the weaknesses of the language ecosystem, 
the founders of the community quickly launched two flagship projects.
Unlike many languages (C, C++ , Python\dots), Fortran did not have a standard library.
The development of a standard \textit{(de facto)} library named stdlib was therefore launched.
One goal is to prevent language users from reinventing the wheel each of them on their own, 
and on the other hand to serve as a laboratory to propose new functionalities to the standardization committee.
Its version 0.2.1 already contains hundreds of utility functions 
(strings, files, operating system integration, unit tests, logging\dots), various algorithms 
(search, sort, fusion\dots) and mathematical functions 
(linear algebra, special functions, Fast Fourier Transform, random numbers, statistics, numerical integration, optimization\dots).

\subsubsection{The fpm package manager}  

Another problem in the Fortran ecosystem was the dispersion of libraries, 
developed independently and using different building tools. 
The Fortran Package Manager fpm solves it \cite{fpm}. 
It is largely inspired by Cargo, the utility of the Rust language serving both as package and build system manager. 
Creating a Hello World project, compiling it and launching it becomes as simple as typing in a terminal:

\begin{lstlisting}
$ fpm new myproject
$ cd myproject
$ fpm run
\end{lstlisting}

The project options are defined in its manifest, a file in TOML format present at its root. 
The fpm package manager manages dependencies, which it downloads from GitHub. 
Associated with the \texttt{git clone} command, testing a Fortran project available on GitHub can be done in thirty seconds. 
Project tests can be started with the \texttt{fpm test} command.
The community indeed promotes current good practices in programming: automatic testing, usage of a version manager, etc.

Fpm, whose version 0.9 was released in June 2023, is now a popular tool
in the community and gradually legacy libraries are modernized using the latest language features and transformed into fpm packages.
As a cherry on top of the cake, fpm is essentially written in Fortran, showing it is also a general purpose language.
This bold choice allows in particular to attract more contributors from the community and promotes the sustainability of the project.

A new flagship project, the \textit{fpm registry} will offer an official repository for fpm packages. 
The user will be able to easily discover, install or publish Fortran packages, as he can do in Python with PyPI.

\subsubsection{The interactive LFortran compiler}  

One of the community founders, Ondřej Čertík\footnote{Now engineer at GSI Technology, 
a manufacturer of SRAM.}, has been developing since 2019 a new open source compiler LFortran \cite{LFortran}, one of whose key features is that it can be used interactively, for example with Jupyter.
It also has several \textit{backends}: LLVM, C, C++, Julia, WASM and x86-64. 
It is not yet in beta version but its development is advancing rapidly.

\subsubsection{Financial support} 

The effort to develop all these tools is carried out partly within the professional framework of some members, 
as part of student members' studies, or as a hobby for others. 
But it is also possible to fund additional efforts.
Thus, since 2021, the Fortran-lang organization participates every year in the 
\textit{Google Summer of Code (GSoC)} where several students paid by Google were working on its projects \cite{GSoC}. 
At the end of 2022, the German organization \textit{Sovereign Tech Fund,} under Federal Ministry for Economic Affairs and Climate Action, gave the organization the equivalent of three full-time jobs to 
work six months on the fpm package manager and the LFortran compiler \cite{STF}. 
It thus recognizes the importance of the language in climate modeling.

\subsection{The FortranCon international conference} 

There have been conferences on High-Performance Computing (HPC) but the community lacked a conference dedicated solely to the Fortran language.
A first \textit{International Fortran Conference (FortranCon)} took place at the University of
Zurich from 2 to 4 July 2020, but virtually because of the COVID pandemic.
It allowed actors and language users to meet and discuss their work and projects.
The second edition, FortranCon 2021, was held again virtually in Zurich on 23 and 24 September 2021,
with typically 70 to 80 participants present simultaneously on the Zoom video conferencing platform.
The videos of these conferences are freely accessible online \cite{FortranCon}.

\section{And what's next?} 

\subsection{Fortran 2023}

The Fortran 2023 standard is expected to be released in fall 2023, almost 70 years after John Backus' letter to his superior.
With many compilers not yet fully implementing Fortran 2018, 
the new standard does not intentionally provide major features but simply 
many improvements in different parts of the language:
new functions for string handling,
in particular to facilitate C to Fortran string interoperability,
trigonometric functions in degrees, longer source lines, new format descriptors,
some improvements for parallel computing, conditional expressions (same syntax as in C), etc.
The \texttt{simple} attribute allows to define \texttt{pure} functions that do not access any
external data except through its arguments.
\texttt{do concurrent} loops are now able to use reduction variables.
All the new features are summarized in the \textit{The New Features of Fortran 202x} document \cite{Reid202x}. 

We could regret that the language does not evolve faster, and that it is necessary 
to wait a long time for certain improvements to come, while newer languages have generally a more rapid development. 
However being a standardized language has its pros and cons, 
and the Fortran Standards Committee pays particular attention to the 
backward compatibility when new features are added.
Thus, compilers incorporating the latest developments are generally still capable of 
compiling the so-called \textit{legacy} codes, that are sometimes 40 or more years old.
These legacy codes are still actively used in large organizations or companies.

As a side note, ISO standards are paid documents. 
But the working document \textit{Fortran 2023 Working Draft} 
is still available on the J3 committee website  \cite{Fortran2023}.
All committee's working drafts are indeed in open access. 

\subsection{Fortran 202Y}

The development of the next standard, voluntarily loosely named Fortran 202Y, is underway. It should notably develop the concept of generic programming, in particular with \textit{templates} as in C++. 
This is certainly the most awaited novelty by Fortran developers, 
as evidenced by the dedicated directory on the GitHub repository \cite{J3GitHub} of the J3 standards committee, where since the end of 2019 anyone can propose and discuss new features for the language.
The LFortran compiler also serves as an experimental platform for the implementation of this major future evolution.

As an important change at the symbolic level, 
we should finally see declared obsolescent the historic default implicit typing!
Among the other innovations whose study is recommended by the J3, we can note the possibility of defining within the code itself the types used by default for each type of data, 
a Fortran preprocessor, and asynchronous task management.

\section{Conclusion}

As languages have been proliferating, Fortran has become a niche language.
To caricature (because it is also a general language), one could say that
it only does one thing, but it does it well: numerical computation\footnote{In the 2000s,
the slogan of the late g95 compiler was \textit{it’s free crunch time}
and GFortran’s one was \textit{free number crunching FORALL!}.}!
That is why it is still widely used in meteorological computations, 
climate modeling, fluid dynamics, computational chemistry, 
materials modeling \cite{FORGE}, nuclear energy, aerospace, and many other areas. 
It is even used in economics and finance \cite{fehr_2018, arnoud2019}. 
It has adapted to all generations and architectures of computer hardware. 
The 2021 Turing Award was moreover awarded to the American Jack Dongarra\footnote{He had worked 
in particular on Fortran libraries such as EISPACK (1974), 
LINPACK, BLAS, LAPACK (1992) and ScaLAPACK (1993), etc. BLAS is also among the ten computer codes with the greatest impact on science \cite{perkel_2021}.}
``for his pioneering contributions to numerical algorithms and libraries that enabled high performance computational software to keep pace with exponential hardware improvements for over four decades.'' \cite{prixturing2021}


In the scientific computing domain, Fortran is now competing\footnote{While hiding in 
the computing libraries of some of them, for example NumPy or SciPy!} with numerous 
languages and softwares: C, C++, Python, Julia, MATLAB, R, etc.
Does labeling it as an old language really make sense? 
The 1956 textbook was 54 pages long, the Fortran 2023 standard is 681 pages long. The language has continuously incorporated new programming paradigms.
It is thus one of the rare standardized languages to integrate parallel computing paradigms \footnote{ Ada 83 was a forerunner in this regard, but Ada's reliability and safety goals take precedence over performance. True to its original impetus, Fortran's approach to parallelism focuses on numerical performance.}. 
Standardized for nearly six decades, its foundations are stable while allowing it to evolve, 
which is a guarantee of durability for the engineer and the researcher \cite{Hinsen2019}.
On the compiler side, not all include the latest standards, and some have
recently disappeared as Lahey and Absoft, but others appear.
Intel's engineers worked five years to completely overhaul their compiler for LLVM.
The open source Flang compiler is also being redesigned to integrate LLVM,
driven in particular by NVIDIA.
And another new open source compiler already mentioned, LFortran, is in development.

\begin{table}[]
  \scriptsize
	\begin{tabular}{|p{1.6cm}|p{4.2cm}|p{4.8cm}|}
	\hline
	\textbf{Name} & \textbf{URL} & \textbf{Characteristics} \\ \hline
	Fortran PlayGround & \url{https://play.fortran-lang.org/} & From Fortran-lang community; GFortran compiler; stdlib standard library directly accessible \\ \hline
	Compiler Explorer & \url{https://godbolt.org/} & Multi-language; possibility to compare execution results and generated assembly code generated by different compilers; sharing option\dots \\ \hline
	tutorialspoint & \url{https://www.tutorialspoint.com/} & Multi-language with didactic material; GNU compiler; possibility to publicly share the code \\ \hline
  LFortran & \url{https://dev.lfortran.org/} & Compiler in alpha version; graphical display; internal compiler representations and C++ generated code \\ \hline
	\end{tabular}
	\caption{Some online Fortran compilers}
	\label{compil_online}
\end{table}
\normalsize

Finally, Fortran now has a dynamic and organized community, full of projects 
to improve its ecosystem and seasoned with current good practices of software development. 
Ten years after coming out of the top 20 on the TIOBE index \cite{tiobe} of 
most ``popular'' programming languages, Fortran has been returning periodically 
since April 2021, even reaching 11{\it th} place in July 2023. 
Interested readers will find online some good French courses \cite{CoursIDRIS, CoursLefrere} 
to approach learning the language or learn its latest features, at least up to Fortran 2008. 
For Fortran 2018, it will be necessary for the moment to refer to the English books quoted previously 
\cite{curcic_modern_2020, metcalf_modern_2018}. 
It is also possible to learn modern Fortran using one of the online compilers listed in Table \ref{compil_online}.

\section{Acknowledgements}

We thank Pierre Hugonnet and Jeremie Vandenplas for their manuscript review and their precious remarks, 
as well as all the members of the Fortran-lang community for the rich and high-quality discussions we 
have on the Fortran Discourse, for the vision of its founders and for the efforts made to improve the language ecosystem.

\bibliographystyle{plain}

\bibliography{paper}

\begin{thebibliography}{10}

\bibitem{TOP500}
{Earth Simulator}:
  \url{https://www.top500.org/resources/top-systems/the-earth-simulator-earth-simulator-center/}.

\bibitem{manuel1956}
{\em Fortran - Programmer's reference manual:
  \emph{\url{http://archive.computerhistory.org/resources/text/Fortran/102649787.05.01.acc.pdf}}}.

\bibitem{Fortran2023}
{Fortran} 2023 standard draft:
  \url{https://j3-fortran.org/doc/year/23/23-007r1.pdf}.

\bibitem{STF}
Fortran at the {Sovereign Tech Fund}:
  \url{https://sovereigntechfund.de/fortran_en}.

\bibitem{GSoC}
Fortran-lang at the {Google Summer of Code}:
  \url{https://summerofcode.withgoogle.com/programs/2023/organizations/fortran-lang}.

\bibitem{fpm}
{Fortran Package Manager} fpm: \url{https://github.com/fortran-lang/fpm}.

\bibitem{FortranCon}
{FortranCon} videos: \url{https://www.youtube.com/@fortrancon6189}.

\bibitem{CoursIDRIS}
The {IDRIS} ({CNRS Institute for Development and Resources in Intensive
  Scientific Computing}) {Fortran} course in {French}:
  \url{http://www.idris.fr/formations/fortran/}.

\bibitem{J3GitHub}
{J3} committee repository: \url{https://github.com/j3-fortran}.

\bibitem{LFortran}
{LFortran} compiler: \url{https://lfortran.org/}.

\bibitem{prototype}
{Prototype-based programming}:
  \url{https://en.wikipedia.org/wiki/Prototype-based_programming}.

\bibitem{PVM}
{PVM (Parallel Virtual Machine)}: \url{https://www.csm.ornl.gov/pvm/}.

\bibitem{FORGE}
Software {FORGE\textregistered} for simulation of hot and cold forming
  processes: \url{https://www.transvalor.com/en/forge}.

\bibitem{tiobe}
{TIOBE} index (most popular languages):
  \url{https://www.tiobe.com/tiobe-index/fortran/}.

\bibitem{CoursLefrere}
Introduction au fortran 90/95/2003/2008 : polycopié de {Master de Jacques
  Lefrère, Université Pierre et Marie Curie Sorbonne Université} :
  \url{http://wwwens.aero.jussieu.fr/lefrere/master/mni/f90+c/polyf90.pdf},
  November 2019.

\bibitem{prixturing2006}
Frances ("Fran")~Elizabeth Allen.
\newblock {A.M. Turing Award} 2006:
  \url{https://amturing.acm.org/award_winners/allen_1012327.cfm}.

\bibitem{arnoud2019}
Antoine Arnoud, Fatih Guvenen, and Tatjana Kleineberg.
\newblock Benchmarking {Global} {Optimizers}.
\newblock Technical Report w26340, National Bureau of Economic Research,
  Cambridge, MA, October 2019.

\bibitem{Backus1957}
J.~W. Backus, R.~J. Beeber, S.~Best, R.~Goldberg, L.~M. Haibt, H.~L. Herrick,
  R.~A. Nelson, D.~Sayre, P.~B. Sheridan, H.~Stern, I.~Ziller, R.~A. Hughes,
  and R.~Nutt.
\newblock The {FORTRAN Automatic Coding System}.
\newblock In {\em Papers Presented at the February 26-28, 1957, Western Joint
  Computer Conference: Techniques for Reliability}, IRE-AIEE-ACM '57 (Western),
  page 188–198, New York, NY, USA, 1957. Association for Computing Machinery.

\bibitem{prixturing1977}
John Backus.
\newblock {A.M. Turing Award} 1977:
  \url{https://amturing.acm.org/award_winners/backus_0703524.cfm}.

\bibitem{curcic_modern_2020}
Milan Curcic.
\newblock {\em Modern {Fortran}: {Building} efficient parallel applications}.
\newblock Manning Publications, first edition, November 2020.

\bibitem{prixturing2021}
Jack Dongarra.
\newblock {A.M. Turing Award} 2021:
  \url{https://amturing.acm.org/award_winners/dongarra_3406337.cfm}.

\bibitem{fehr_2018}
Hans Fehr and Fabian Kindermann.
\newblock {\em Introduction to {Computational} {Economics} {Using} {Fortran}}.
\newblock Oxford University Press, May 2018.

\bibitem{marytsingou}
Virginia Grant.
\newblock {We thank Miss Mary Tsingou}, December 2020.

\bibitem{heising_emergence_1984}
W.~P. Heising.
\newblock The {Emergence} of {FORTRAN} {IV} from {FORTRAN} {II}.
\newblock In {\em Annals of the {History} of {Computing}}, volume~1, pages
  28--32, January 1984.

\bibitem{Hinsen2019}
K.~{Hinsen}.
\newblock {Dealing With Software Collapse}.
\newblock {\em Computing in Science Engineering}, 21(3):104--108, 2019.

\bibitem{kedward_state_2022}
Laurence~J. Kedward, Balint Aradi, Ondrej Certik, Milan Curcic, Sebastian
  Ehlert, Philipp Engel, Rohit Goswami, Michael Hirsch, Asdrubal Lozada-Blanco,
  Vincent Magnin, Arjen Markus, Emanuele Pagone, Ivan Pribec, Brad Richardson,
  Harris Snyder, John Urban, and Jeremie Vandenplas.
\newblock The {State} of {Fortran}.
\newblock {\em Computing in Science \& Engineering}, 24(2):63--72, March 2022.

\bibitem{kennedy2007}
Ken Kennedy, Charles Koelbel, and Hans Zima.
\newblock The rise and fall of {High} {Performance} {Fortran}: an historical
  object lesson.
\newblock In {\em Proceedings of the third {ACM} {SIGPLAN} conference on
  {History} of programming languages}, San Diego California, June 2007. ACM.

\bibitem{lorenzo_abstracting_2019}
Mark~Jones Lorenzo.
\newblock {\em Abstracting {Away} the {Machine}: {The} {History} of the
  {FORTRAN} {Programming} {Language} ({FORmula} {TRANslation})}.
\newblock SE Books, August 2019.

\bibitem{gtk-fortran2019}
Vincent Magnin, James Tappin, Jens Hunger, and Jerry DeLisle.
\newblock gtk-fortran: a {GTK}+ binding to build {Graphical} {User}
  {Interfaces} in {Fortran}.
\newblock {\em The Journal of Open Source Software}, February 2019.

\bibitem{calculparallele}
Frédéric Magoulès and François-Xavier Roux.
\newblock {\em Calcul scientifique parallèle}.
\newblock October 2017.

\bibitem{markus2012}
Arjen Markus.
\newblock {\em Modern {Fortran} in {Practice}}.
\newblock Cambridge University Press, Cambridge, 2012.

\bibitem{metcalf_modern_2018}
Michael Metcalf, John~Ker Reid, and Malcolm Cohen.
\newblock {\em Modern {Fortran} explained: incorporating {Fortran} 2018}.
\newblock Numerical mathematics and scientific computation. Oxford University
  Press, 2018.

\bibitem{numrich2005}
Robert~W. Numrich.
\newblock Parallel numerical algorithms based on tensor notation and
  {Co}-{Array} {Fortran} syntax.
\newblock {\em Parallel Computing}, 31(6):588--607, June 2005.

\bibitem{perkel_2021}
Jeffrey~M. Perkel.
\newblock Ten computer codes that transformed science.
\newblock {\em Nature}, 589(7842):344--348, January 2021.

\bibitem{Reid202x}
John Reid.
\newblock The {New Features of Fortran} 202x.
  \url{https://wg5-fortran.org/N2151-N2200/N2194.pdf}.
\newblock Technical report, 2022.

\bibitem{rouson2014}
Damian Rouson, Jim Xia, and Xiaofeng Xu.
\newblock {\em Scientific software design: the object-oriented way}.
\newblock Cambridge Univ. Press, Cambridge, 1. pbk. ed edition, 2014.

\end{thebibliography}

\end{document}